\documentclass[reprint,amsmath,amssymb,pre,superscriptaddress]{revtex4-1}
\usepackage{graphicx}
\usepackage{dcolumn}
\usepackage{bm}
\usepackage[utf8]{inputenc}

\begin{document}

\title{Analysis of data in the form of graphs}

\author{Karthikeyan Rajendran}
\email{krajendr@princeton.edu}
\homepage{http://arnold.princeton.edu/~krajendr/}
\affiliation{Department of Chemical and Biological Engineering, Princeton University, Princeton, New Jersey 08544, USA}
\author{Ioannis G. Kevrekidis}
\email{yannis@princeton.edu}
\homepage{http://arnold.princeton.edu/~yannis/}
\affiliation{Department of Chemical and Biological Engineering, Princeton University, Princeton, New Jersey 08544, USA}
\affiliation{Program in Applied and Computational Mathematics (PACM), Princeton University, Princeton, New Jersey 08544, USA}

\date{\today}
\begin{abstract}
We discuss the problem of extending data mining approaches to
cases in which data points arise in the form of individual graphs.
Being able to find the intrinsic low-dimensionality in ensembles of
graphs can be useful in a variety of modeling contexts, especially
when coarse-graining the detailed graph information is of interest.
One of the main challenges in mining graph data is the definition
of a suitable pairwise similarity metric in the space of graphs.
We explore two practical solutions to solving this problem: one based on
finding subgraph densities, and one using spectral information.
The approach is illustrated on three test data sets (ensembles of graphs);
two of these are obtained from standard graph generating algorithms,
while the graphs in the third example are sampled as dynamic snapshots
from an evolving network simulation.
\end{abstract}


\maketitle

\section{\label{sec:intro}Introduction}

Recent advances in scientific computation (both hardware and algorithms)
increasingly facilitate the direct simulation of complex systems, characterized by
very large numbers of degrees of freedom;
detailed dynamic simulations are extensively used beyond statistical mechanics,
across fields such as epidemiology \cite{Euba04modelling,Ferg05strategies,Long86predicting},
economics \cite{Iori02microsimulation,Wang05microscopic} or
biology \cite{Levi01self-organization,Liu04stable}.
Large-scale graphs -networks- constitute a key feature of many
complex systems \cite{Bara02linked:,Newm03structure}, whether in the form
of a large {\em fixed} network of biochemical reactions, or in the form
of an {\em evolving} social network.
Tools for the systematic analysis and characterization of networks, and, in particular,
tools for the concise (coarse-grained) description of their state are crucial
in modeling dynamic behavior, especially when the networks become very large.
{\em A priori} knowledge of the right few observables, that is, the
(hopefully) small number of network statistics  that suffice to summarize the
salient network features, is clearly a major advantage for the modeler.
When such observables (``coarse variables") are not known, their detection from
data-mining becomes an important and challenging task, since standard data mining approaches
need to be extended to cases where each data point is in the form of a graph.
Such data ensembles would arise from recording many ``snapshots" of a dynamic network
evolution process, such as a growing social network, or from the action of plasticity
in a neuronal network; they might also arise from ``horizontally" (rather than longitudinally)
sampling the social networks of several city residents at the same moment in time, or
the contact networks arising in several granular experiments.

Good low-dimensional observables for an ensemble of graph data can find
several uses in a modeling context.
One can, for example, try to close effective dynamic equations in terms of
these reduced observables, explicitly if possible through appropriate
assumptions, or even ``on the fly" numerically, in the spirit of
multiscale modeling algorithms like those of the equation-free approach
(see \cite{Kevr03Equation-free,Kevr04Equation-free:}
and for applications to network problems \cite{Bold12Equation-Free,Tsou12coarse-graining}).

Low-dimensionality in the collected data ensemble also has implications
about separation of time scales and long-term dynamics for graph evolution
processes (e.g. number of slow eigenvalues in the final approach to
a stable stationary state, or exploration of coarse-grained instabilities
and bifurcations as model parameters vary).
In the case of graph-generating algorithms (like \cite{Watt98collective,Albe02statistical}),
it should be useful to compare the dimensionality of the resulting ensemble to the number of free parameters
in the model, possibly leading to model fine-tuning modifications.
A particularly interesting {\em inverse} problem also arises: given values of
a few important observables, when -and how- is it possible to construct actual network
realizations consistent with these values?
Graph generating algorithms motivated by such inverse problems can be found
in the literature, both deterministic (e.g. the  Havel-Hakimi algorithm for construction
of networks with a prescribed degree sequence \cite{Have55remark})
and stochastic (e.g. \cite{Doro02how,Serr05tuning}).
A recent mixed integer linear programming formulation of such inverse problems
can be found in \cite{Goun11generation}.
In this paper we will concentrate on determining reduced sets of observables (good reduction coordinates)
from graph ensemble data.
There is, of course, no guarantee that these data mining based parameterizations of graph data
have a simple and obvious associated physical meaning; if simple and easily communicable meaning
is desirable, then additional effort has to be invested in finding sets of physically meaningful
variables that are one-to-one (on the graph data) with the ones discovered through data mining.

The paper is organized as follows:
In Section \ref{sec:dm}, we briefly discuss the data mining algorithm
that will be used in this paper.
Subsequently, in Section \ref{sec:sim}, we focus on the issue of defining pairwise similarities between graphs;
this constitutes the biggest challenge in adapting established data mining techniques to
data in the form of graphs.
We discuss the two options we use in approaching this problem.
Three illustrative network ensemble examples are presented,
and data mining algorithms (with both our
similarity measure choices) are implemented in Section ~\ref{sec:res}.
A summary of results and suggestions for future work are finally presented in
Section ~\ref{sec:conc}.

\section{\label{sec:dm}Data mining}

The most established tool in data mining is, arguably, principal component analysis (PCA) \cite{Shle03tutorial},
which is used to represent a low dimensional dataset, embedded in a (much) higher dimensional space,
in terms of an optimal (in a well-defined sense) linear basis.
It enables one to identify directions along which the data ensemble has the most variance.
But PCA can only find out the best {\em linear} lower
dimensional subspace in which the dataset lies.
In many problems, however, the data lie in a highly curved, non-linear
lower dimensional subspace; the best linear subspace required to embed the
data may thus be significantly higher-dimensional than the inherent data dimensionality.
A number of non-linear data mining tools such as Diffusion Maps \cite{Nadl05diffusion,Nadl06diffusion}
and ISOMAP \cite{Tene00global} have been developed with the goal to extract such an
inherently nonlinear minimal-dimensional subspace intrinsic to the data.
We use diffusion maps as a representative non-linear data mining approach,
modified so as to apply to sets of data in the form of graphs.
In diffusion maps one constructs a graph with the data points as vertices,
while a pairwise similarity measure between two data points is used as the weight on the corresponding edge.
In broad terms, the leading eigenfunctions of the diffusion process on this graph are used to embed the data points.
If the data points actually lie very close to (in) a low dimensional non-linear manifold,
the first few of these eigenfunctions will suffice to embed the data while retaining most(all) the
relevant information.

Consider a set of {\em n} points $\{x_i\}_{i=1}^n$ in {\em p}-dimensional {\em Euclidean} space, one
defines a similarity matrix, $W$ (a measure of closeness between pairs of
points in this space) according to the following equation (choosing a Gaussian kernel):

\begin{equation}
W(i,j) = exp \left( \frac{-\|x_i-x_j\|^2}{\epsilon^2} \right).
\label{eq:sim}
\end{equation}

Here, $\epsilon$ is a suitable length scale; this is a tuning parameter:
points with Euclidean distance (much) larger than $\epsilon$ are,
effectively, not connected directly.
One also defines a diagonal normalization matrix, $D_{ii}=\sum_j W_{ij}$,
and consequently the matrix, $A=D^{-1} W$.
$A$ can be viewed as a Markov matrix defining a random walk (or diffusion) on the data points,
i.e., $A_{ij}$ denotes the probability of transition from  $x_i$ to $x_j$.
Since $A$ is a Markov matrix, the first eigenvalue is always $1$.
The corresponding eigenvector is a constant, trivial eigenvector.
In diffusion maps, the next few non-trivial eigenvectors of $A$
(corresponding to the next few largest eigenvalues) constitute the best ``directions"
that parameterize the nonlinear manifold in which the data lie.

It is clear that a crucial step in the implementation
of diffusion maps is the definition of a measure of pairwise similarity between data points.
If the data points are defined in a Euclidean space, it is straightforward to use the
Euclidean norm as the  measure the distance (the closeness) between pairs of points.
When the data points are given in the form of individual graphs, however, it is not
trivial to define good measures of pairwise similarity between them.
Thus, the important step in successfully adapting the machinery of
non-linear data mining -and diffusion maps in particular- to the case of graph-type
data lies in our ability to define a ``useful" measure of similarity/closeness between pairs of graphs.

\section{\label{sec:sim} Defining similarity measures between graph objects}

Although measures of similarity in the context of graphs have been discussed in the
literature \cite{Dana11algorithms}, standardized classifications
are still lacking.
One may define similarities between nodes in a given graph,
or similarities between the graphs themselves.
In this paper, we will discuss the latter type,
since we are interested in comparing entire graph objects.
Additionally, the nodes of the graphs may be labeled or unlabeled, giving
rise to different similarity measures.
We are interested in the case of unlabeled nodes,
where the problem of ``matching" the nodes across the two graphs
(i.e. ordering the nodes) makes the definition of similarity
measures more challenging.
We will focus here on the case where all the graphs in the dataset have
{\em the same number of nodes}; the approach is, in principle, easily
extendable to collections of graphs of different sizes.

Existing techniques in the literature for defining graph similarities
may roughly be classified into a few broad categories.
The first is the class of methods that make use of
the {\em structure} of the graphs to define similarities.
An obvious choice is to consider two graphs to be similar
if they are isomorphic \cite{Peli98replicator}.
One of the first definitions of distance between pairs of graphs
using the idea of graph isomorphism was based on constructing the
smallest larger graph with subsets isomorphic to both graphs of interest \cite{Zeli75certain}.
Likewise, one can define similarity measures based on the largest
common subgraph in pairs of graphs \cite{Bunk98graph,Raym02rascal:}.
The {\em graph edit distance}, which measures the number of operations on the nodes
and edges of a graph required to transform it into the other graph, is another
example of a method using the idea of graph isomorphism.
The graph edit distance and a list of other measures that use the
structure of the network to quantify similarity are discussed in \cite{Papa08web}.

One can alternatively use methods that compare the behavior of the neighborhoods of
the nodes in the two graphs.
Comparing neighborhoods of nodes is especially applicable in measuring similarities
between sparse graphs.
Often, the graph similarity problem is solved through solving the related
problem of graph matching, which entails finding the correspondence between
nodes in the two graphs such that the edge overlap is maximal.
Methods like the similarity flooding algorithm \cite{Meln02similarity},
the graph similarity scoring algorithm \cite{Zage08graph} and the belief
propagation algorithm \cite{Bay09message} are representative of this type of approach.
Graph kernels based on the idea of random walks \cite{Gaert03graph,Kash03marginalized,Mahe04extensions}
also fall under the category of algorithms based on comparing node neighborhoods.

One of the simplest options to evaluate similarities between graphs is to
directly compare a few chosen, representative features of the network.
The chosen features may correspond to any facet of the graph, such as
structural information (degree distribution, for instance) or spectral measures
(eigenvalues and/or eigenvectors of the graph Laplacian matrix).
In this paper, we will take this approach and consider two options
for defining similarities between graphs:
(i) using a list of several subgraph densities and (ii) an approach using spectral information.

\subsection{\label{ss:den} The subgraph density approach}

The general idea behind this approach is that two graphs
are similar if {\em the frequencies of occurrence of representative subgraphs
in these graphs are similar.}
The density of a small subgraph in a large graph is a weighted
frequency of occurrence of the subgraph (pattern) in the large, original graph.
We use the following definition for the density of a subgraph $H$ with $k$ nodes
in a graph $G$ with $n$ nodes:

\begin{equation}
\label{eqn:homdenG}
\rho(H,G) := \frac{1}{{n\choose k}} \!  \sum_{\varphi:[k] \to [n]} \!
\left[ \forall  i, j \in [k] \! : \! H(i,j) \! = \! G_n(\varphi(i),
\varphi(j)) \right].
\end{equation}

A graph can be reconstructed exactly if the densities of {\em all possible}
subgraphs in the graph are specified \cite{Lov06limits}.
Thus, a list of all these subgraph densities is an alternative way
of providing complete information about a graph.
This list can be thought of as an embedding of the graph, which can
then be used to define similarity measures in the space of graphs.
It is, however, not so practical to computationally find the subgraph densities of {\em all}
possible subgraphs of a given graph, especially when the number of
nodes in the graph becomes large.
A systematic yet practical way to embed a graph is to use the subgraph densities
of all subgraphs lesser than a given (relatively small, and thus computationally
manageable) size in the graph.
For instance, to embed a graph with $n$ nodes, one can evaluate the subgraph densities
of all subgraphs of size less than or equal to $m~(m \ll n)$.
Since $m \ll n$, the embedding cannot be used to {\em exactly} reconstruct the graph.
One can, nevertheless, compute distances between the embeddings
(the vectors of subgraph densities) of any two graphs; these distances
may be useful estimates of the similarities between the graphs.

Let $G_i$ and $G_j$ be two graphs defined on $n$ nodes.
Let $H_1$, $H_2$,$\ldots H_r$ be the $r$ chosen, representative subgraphs.
We find the frequencies of occurrence of these subgraphs in the original
graphs by appropriately modifying the open-source RANDESU algorithm described
in \cite{Wern06fanmod:}.
The subgraph densities are calculated (normalized) by dividing these frequencies by
$n\choose{k}$, where $n$ and $k$ are the the number of nodes in the
original graph and the subgraph respectively.
Although dividing by $n\choose{k}$ is not a unique choice
for normalizing the subgraph densities, the densities we calculated this way had similar
orders of magnitude; we thus considered them a sensible choice.
The density of subgraph $H$ in graph $G$ is denoted by $\rho(H,G))$.
The similarity measure between a pair of graphs $G_i$ and $G_j$ can then
be defined as an $L_2$-norm (possibly weighted) of the difference between the vectors of subgraph densities
as follows:

\begin{equation}
k(G_i,G_j) = \sqrt{\sum_{l=1}^{r}\left(\rho(H_l,G_i)-\rho(H_l,G_j)\right)^2}.
\label{eq:k1}
\end{equation}

In order to use this pairwise similarity measure in a diffusion map context, the
Gaussian kernel, analogous to Eq.~\ref{eq:sim}, can be calculated as follows:

\begin{equation}
W(i,j) = exp \left( \frac{-(k(G_i,G_j))^2}{\epsilon^2} \right).
\label{eq:sim1}
\end{equation}

\begin{figure}
\begin{center}
\includegraphics[width=0.36\textwidth]{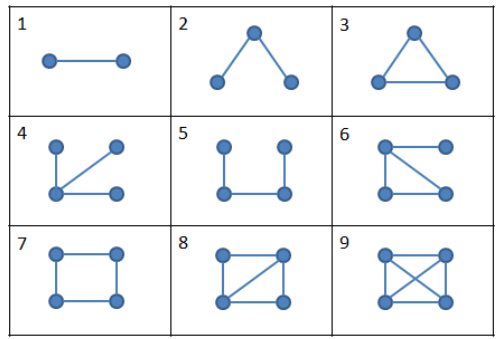}
\caption{\label{fig:SGs} The $9$ connected subgraphs of size less than or equal to $4$.
}
\end{center}
\end{figure}

In our illustrative numerical computations, we considered all connected subgraphs of size less than
or equal to $m=4$ as a representative sample of subgraphs.
There are $r=9$ such graphs as shown in Fig.~\ref{fig:SGs}.

\subsection{\label{ss:s} Our spectral approach}

Our second approach to defining similarities between graphs was initially motivated by the approach given
in \cite{Vis10graph}, and we based it on the notion of non-conservative diffusion on graphs \cite{Gho11non-conservative}.
There are, of course, numerous ways in which spectral
information of graphs (or equivalently information from performing random walks on graphs)
could be used to define similarity measures.
The particular version of the similarity metric discussed here is inspired by the
spectral decomposition algorithm in \cite{Vis10graph}.
The usual definition of random walks on graphs is based on a physical diffusion process.
One starts with a given initial density of random walkers, who are then redistributed at
every step by premultiplying the distribution of random walkers at the current stage by
the graph adjacency matrix.
The rows of the adjacency matrix are scaled by the row sum,
so that the quantity of random walkers is conserved.
In our approach, we consider a non-conservative diffusion process where we replace the
normalized adjacency matrix in the random walk process by its original, unnormalized counterpart.

Let us consider graphs $G_i$ and $G_j$, with adjacency matrices $B_i$ and $B_j$ respectively.
Let their spectral decompositions be given by $B_i = P_i D_i P_i^{T}$ and $B_j = P_j D_j P_j^{T}$ respectively.
Let the initial probability distribution of random walkers on the $n$ nodes of the graph be denoted by $\hat{p}$.
This can be taken to be the uniform distribution.
At every step of the process, the new distribution of random walkers is found
by applying the unnormalized adjacency matrix to the distribution at the previous step.
Since the adjacency matrix is not normalized, the density of random walkers
changes over time depending on the weights associated with the edges of the graphs.
We consider walks of different lengths, at the end of which we evaluate
statistics by weighing the density of random walkers on the nodes according
to a prescribed vector $\hat{q}$; this can also be assumed to be a uniform vector
that takes the value $1/n$ at every node.
As pointed out in \cite{Vis10graph}, the vectors $\hat{p}$ and $\hat{q}$
are ways to \emph{``embed prior knowledge into the kernel design"}.
We reiterate that, although the method is general, we will consider the special case
where the sizes of the graphs are the same.

The (possibly weighted) average density of random walkers after a {\em k}-length walk in $G_i$, is denoted by $Q_{ik}$.
This can be evaluated as follows:

\begin{equation}
Q_{ik} = \hat{q}^{T}B_i^k \hat{p} = \hat{q}^{T} (P_i D_i^k P_i^{T}) \hat{p}.
\end{equation}

Consider a summation of $Q_{ik}$ for walks of all lengths, with appropriate weights $\mu_k$
corresponding to each value of $k$.
Let the computed weighted sum of densities corresponding to graph $G_i$ be denoted as $S_i$.

\begin{equation}
S_i = \sum_{k=0}^{\infty} \mu(k) Q_{ik} = \sum_{k=0}^{\infty} \mu(k)~l_i^{T} D_i^k r_i.
\end{equation}

where $l_i=P_i^{T}\hat{q}$ and $r_i=P_i^{T} \hat{p}$.

We used the following choice of weighting relation: $\mu(k) = \frac{\lambda^k}{k!}$.
With this choice of weights, one can write $S_i$ as a simple function of $\lambda$ as follows:

\begin{equation}
S_i(\lambda) = l_i^{T}e^{(\lambda D_i)}r_i.
\label{eq:S}
\end{equation}

Thus, every graph $G_i$ is embedded using these $S_i$ values evaluated at characteristic
values of $\lambda$ (say $\lambda_1, \lambda_2,... \lambda_M$)\footnote{Note that
an alternative equivalent way to define the similarity measure would be to directly
compare the contribution of the different eigenvectors to $S_i$ instead of
summing the contributions and then using different values of $\lambda$.
However, it is not so straightforward to generalize this alternative approach
to cases where there are graph sizes vary within the dataset.}.
A similarity between any two graphs $G_i$ and $G_j$ can then be evaluated using the
Gaussian kernel defined in Eq.~\ref{eq:sim1} using the following expression for $k(G_i,G_j)$:

\begin{equation}
k(G_i,G_j) = \sqrt{\sum_{m=1}^{M} (S_i(\lambda_m)-S_j(\lambda_m))^2}.
\label{eq:k2}
\end{equation}

This formula is simple and convenient for our purpose.
For every graph $G_i$, one can evaluate the three vectors,
$l_i$, diagonal elements of $D_i$ and $r_i$ and store them.
The $3n$ values they are comprised of can be thought of (and used)
as a coarse embedding of the graph.
The similarity measure between pairs of graphs can then finally be
evaluated by using Eq.~\ref{eq:S} and Eq.~\ref{eq:k2}
in terms of these stored values.
This also makes it straightforward to add new graphs in the ensemble, increasing the size of
the similarity matrix without much additional computation.

\section{\label{sec:res} Computational results}

We will explore the dimensionality of datasets (where the data points are individual
graphs) using the diffusion map approach; within this approach we will construct
implementations using the graph similarity metrics mentioned above.
We use three different datasets for this exploration; two of them arise in the
context of ``graph-generation" models (they are the ubiquitous Erd\"{o}s-R\'{e}nyi
networks and the Chung-Lu networks).
The third is closer to the types of applications
that motivated our work: networks that arise as individual temporal ``snapshots" during a
dynamic network evolution problem.
%
%

\subsection{\label{ss:er} Test case $1$: Erd\"{o}s-R\'{e}nyi graphs}

\begin{figure*}
\begin{center}
\includegraphics[width=0.81\textwidth]{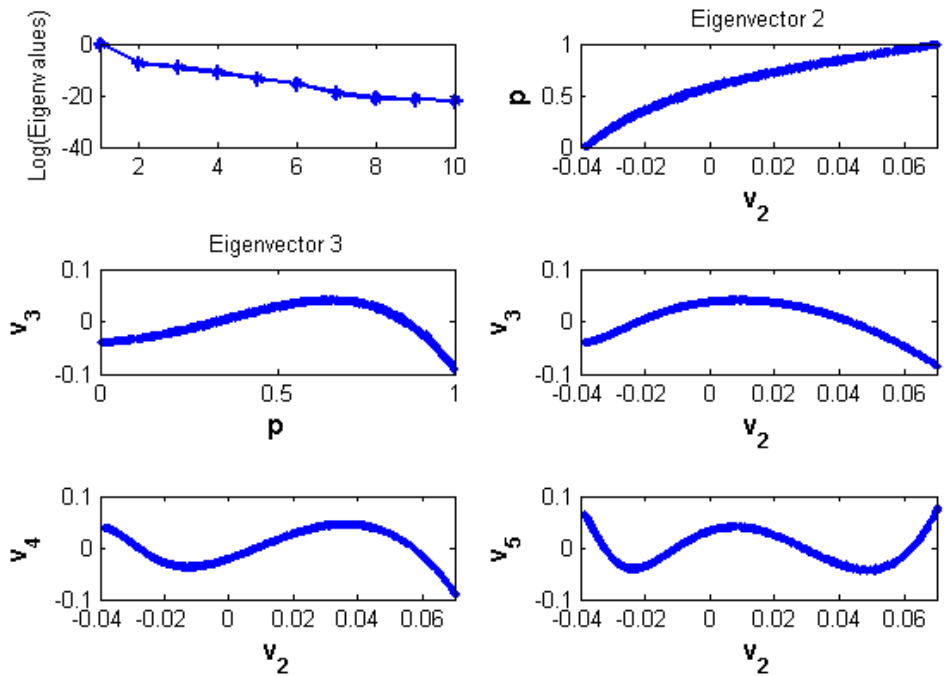}
\caption{\label{fig:ER1} Data mining ensembles of Erd\"{o}s-R\'{e}nyi graphs: The
subgraph approach was used to quantify similarity between individual graphs.
The top-left plot shows the first $10$ eigenvalues of the random walk matrix arising in Diffusion Maps.
The corresponding first two non-trivial eigenvectors are plotted against the ``construction parameter" $p$
used to create the graphs, as well as against each other.
Notice how the first non-trivial eigenvector (the second eigenvector) is one-to-one with $p$.
The fourth and fifth eigenvectors are also plotted against the second eigenvector.
An $\epsilon$ of 0.5 was used in the diffusion map algorithm.
}
\end{center}
\end{figure*}

\begin{figure*}
\begin{center}
\includegraphics[width=0.81\textwidth]{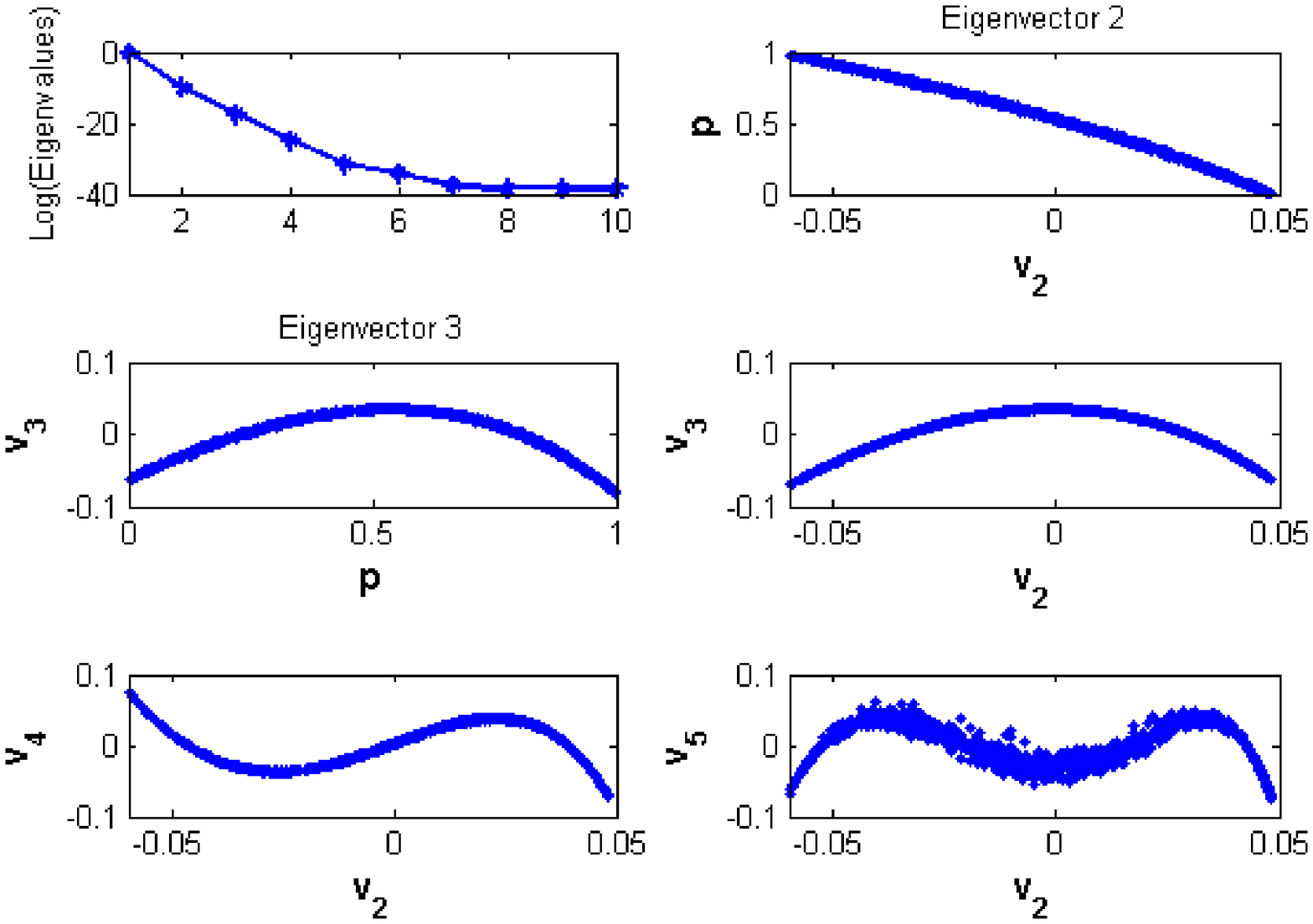}
\caption{\label{fig:ER2} Data mining ensembles of Erd\"{o}s-R\'{e}nyi graphs: Our
spectral approach was used to quantify similarity between graphs. The top-left plot shows
the first $10$ eigenvalues of the random walk matrix arising in Diffusion Maps. The
corresponding first two non-trivial eigenvectors are plotted against the construction parameter $p$
used to create the graphs, as well as against each other.
Notice how, again, the first non-trivial eigenvector (the second eigenvector) is
one-to-one with $p$.
The fourth and fifth eigenvectors are also plotted against the second eigenvector.
An $\epsilon$ of 5 was used in the diffusion map algorithm.
}
\end{center}
\end{figure*}

Consider, as our initial example, a dataset consisting of $m=1000$
Erd\"{o}s-R\'{e}nyi $G(n,p)$ random graphs \cite{Erd59random} with $n=100$ nodes each.
The parameter $p$ (the probability of edge existence) used to construct these
graphs is randomly sampled uniformly in the interval $(0,1)$.
%
%
We start by computing the similarity measures between pairs of individual graphs
(both the subgraph approach (Eq.~\ref{eq:k1}) using $9$ subgraph densities
and our spectral approach (Eq.~\ref{eq:k2} using 100 values of $\lambda$ uniformly spaced from $0.0001$ to $0.01$).
The similarity matrix $W$ is then calculated using Eq.~\ref{eq:sim1}.
The first $10$ eigenvalues of the corresponding random walk matrix $A$,
(as described in Sec.~\ref{sec:dm}) are plotted in Figs.~\ref{fig:ER1} and \ref{fig:ER2},
corresponding to the subgraph approach and to our spectral approach respectively.
For both these cases, the first two non-trivial eigenvectors (viz., the eigenvectors corresponding
to the second and third
eigenvalues) are plotted against the parameter $p$ of the corresponding Erd\"{o}s-R\'{e}nyi graph.
From the figures, it is clear that the second eigenvector is
one-to-one with the parameter $p$, which here is also the edge-density.
Thus, this eigenvector (in both cases) captures the principal direction
of variation in the collection of Erd\"{o}s-R\'{e}nyi graphs.
In other words, our data mining approach independently recovers the
single important parameter $p$ in our sample dataset.
%

%
%

In diffusion maps, the first non-trivial eigenvector always characterizes the principal direction in
the dataset.
Subsequent eigenvectors can represent one of the following:
(i) higher harmonics of the principal direction,
(ii) new directions in the dataset, or (iii) noise (in this case, the variability
of sampling among Erd\"{o}s-R\'{e}nyi graphs of the same $p$).
One can discriminate between these options by plotting the eigenvectors
against each other and looking for correlations.
In this example, when subsequent eigenvectors are plotted against the second, we
clearly observe that they are simply higher harmonics in its ``direction".
The third, fourth and fifth eigenvectors, in both cases, are clearly seen to be a non-monotonic function of $v_2$($p$)
but with an increasing number of ``spatial" oscillations, reminiscent of Sturm-Liouville type problem eigenfunction shapes.
These eigenvectors do not, therefore, capture {\em new directions} in the space of our sample graphs.
%

This simple example serves to illustrate the purpose of using
data mining algorithms on graph data.
In this case, we created a one parameter family of graphs,
characterized by the parameter $p$.
Using only the resulting graph objects, our data mining
approach successfully recovered a characterization of these graphs
equivalent to (one-to-one with) this parameter $p$.
One feature of this one-to-one correspondence between $p$ and the $v_2$ component
of the graphs is worth more discussion: data mining discovers the ``one-dimensionality"
of the data ensemble, but does not {\em explicitly} identify $p$ - a parameterization
that has a direct and obvious physical meaning.
Data mining only provides a parameterization effectively isomorphic to the one by $p$:
to the eye the $p$-$v_2$ function appears continuous and with a continuous inverse.
{\em Providing a physical meaning} for the parameterization discovered (or finding a physically
meaningful parameterization isomorphic to the one discovered) is a completely separate
task, where the modeler has to provide good candidates.
The contribution of the data-mining process is determining
the {\em number} of necessary parameters, and in providing a quantity against
which good candidates can be tested.

\subsection{\label{ss:cl} Test case $2$: A two parameter family of graphs}

\begin{figure*}
\begin{center}
\includegraphics[width=0.81\textwidth]{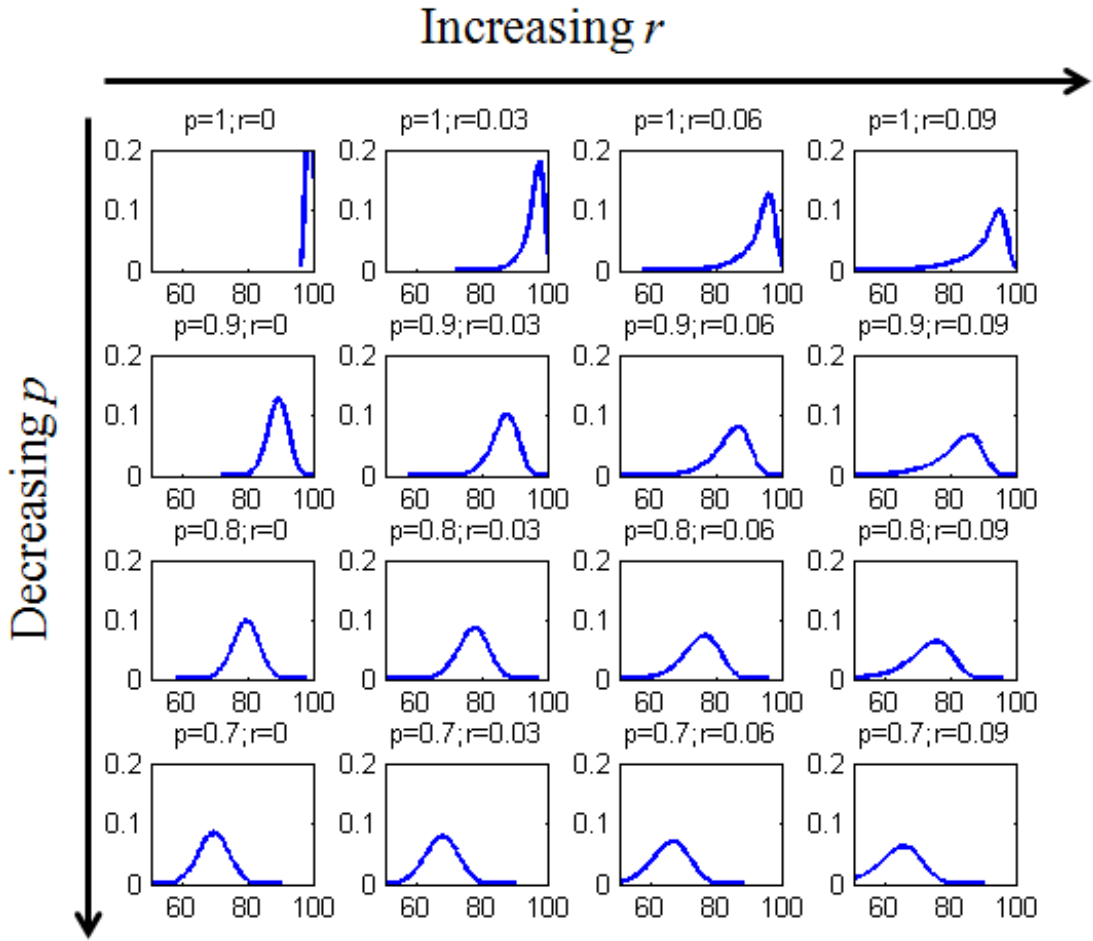}
\caption{\label{fig:CL} The degree distribution of Chung-Lu graphs created using
the algorithm described in the text are plotted for various values of the construction parameters
$p$ and $r$. The parameter $p$ corresponds to the density of edges in the
graph. As $p$ decreases, the degree distribution shifts uniformly to the left.
The parameter $r$ corresponds roughly to the skewness of the degree distribution.
As $r$ is increased from $0$, the degree distribution shifts to the left, but the
resulting degree distributions are skewed more and more to the left.
}
\end{center}
\end{figure*}

\begin{figure*}
\begin{center}
\includegraphics[width=0.81\textwidth]{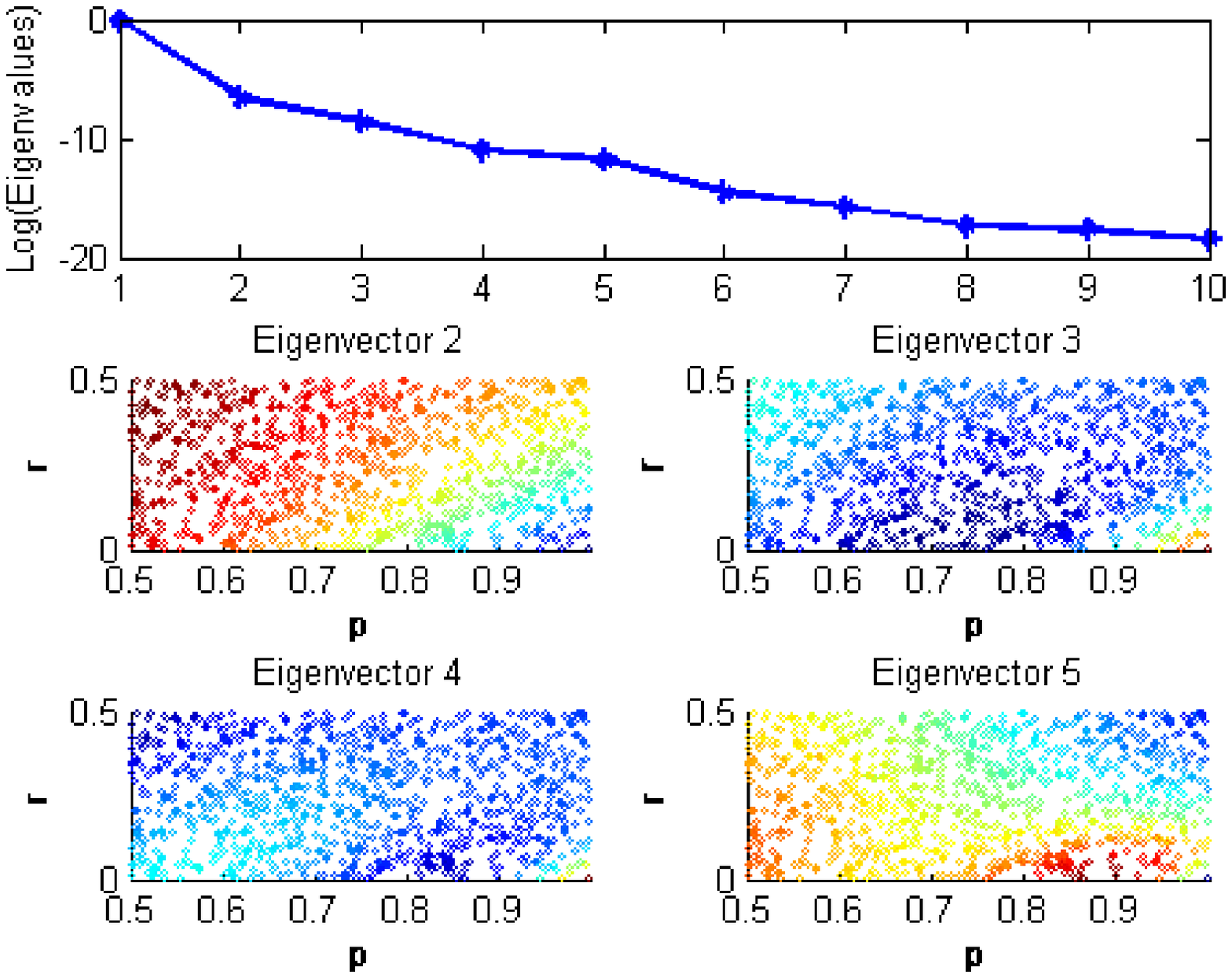}
\caption{\label{fig:CL1} Data mining ensembles of two-parameter Chung-Lu graphs:
The leading eigenvalues of the random walk matrix calculated using the subgraph similarity measure are first plotted.
The corresponding first four non-trivial eigenvectors are then illustrated in a way that brings forth their
relation to the construction parameters $p$ and $r$.
In these plots, each graph is denoted as a point.
The $x$ and $y$ coordinates of the point correspond to the parameters $p$ and $r$ used to construct that particular graph.
The graphs are colored based on the magnitude of their components in the eigenvectors of the random walk matrix $A$.
An $\epsilon$ of 10 was used in the diffusion map algorithm.
}
\end{center}
\end{figure*}

\begin{figure*}
\begin{center}
\includegraphics[width=0.81\textwidth]{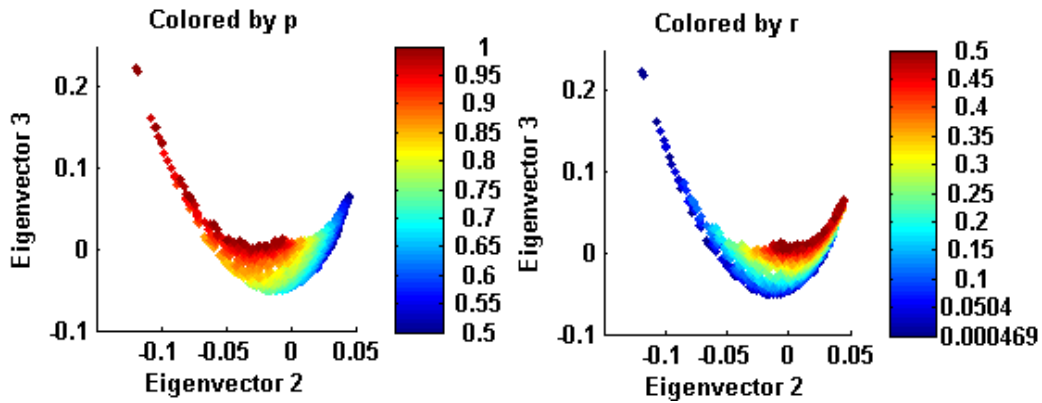}
\caption{\label{fig:CL1a}
Data mining the two-parameter family of Chung-Lu graphs using the subgraph similarity metric
leads to an apparent two-dimensional
embedding.
In these plots the $x$ and $y$ coordinates of each point (i.e. of each graph in the dataset)
denote the components
of that particular graph in the second and third eigenvectors of the random walk matrix respectively.
Each point is now colored
based on the parameter values of $p$ (left) and $r$ (right) used to construct the particular graph.
}
\end{center}
\end{figure*}

\begin{figure*}
\begin{center}
\includegraphics[width=0.81\textwidth]{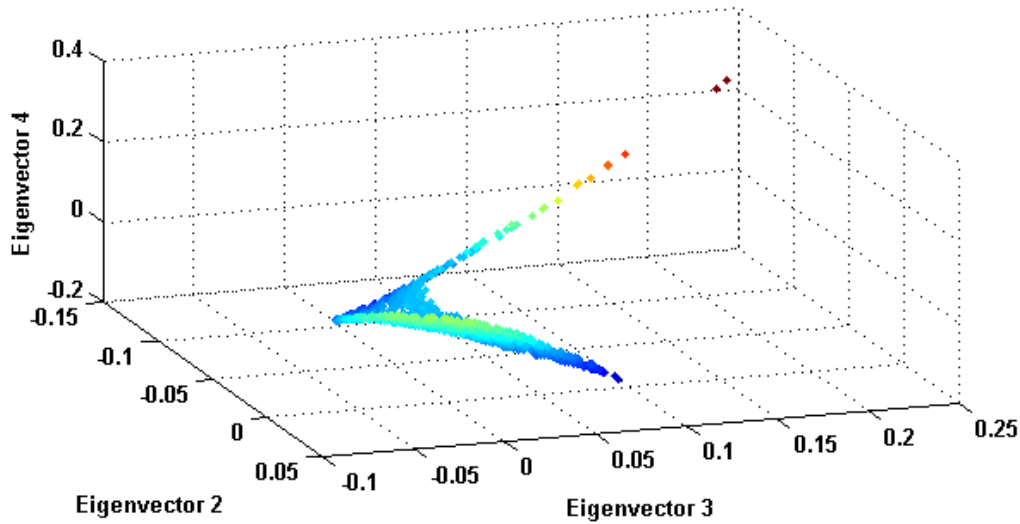}
\caption{\label{fig:CL1b} A $3-d$ plot suggesting that the fourth eigenvector of
the random walk matrix -calculated using the subgraph similarity measure- for the case of the two parameter family of
Chung-Lu graphs
can be expressed as a function of the second and third
eigenvectors: it does not capture a new direction in the space of our sample graphs.
}
\end{center}
\end{figure*}

\begin{figure*}
\begin{center}
\includegraphics[width=0.81\textwidth]{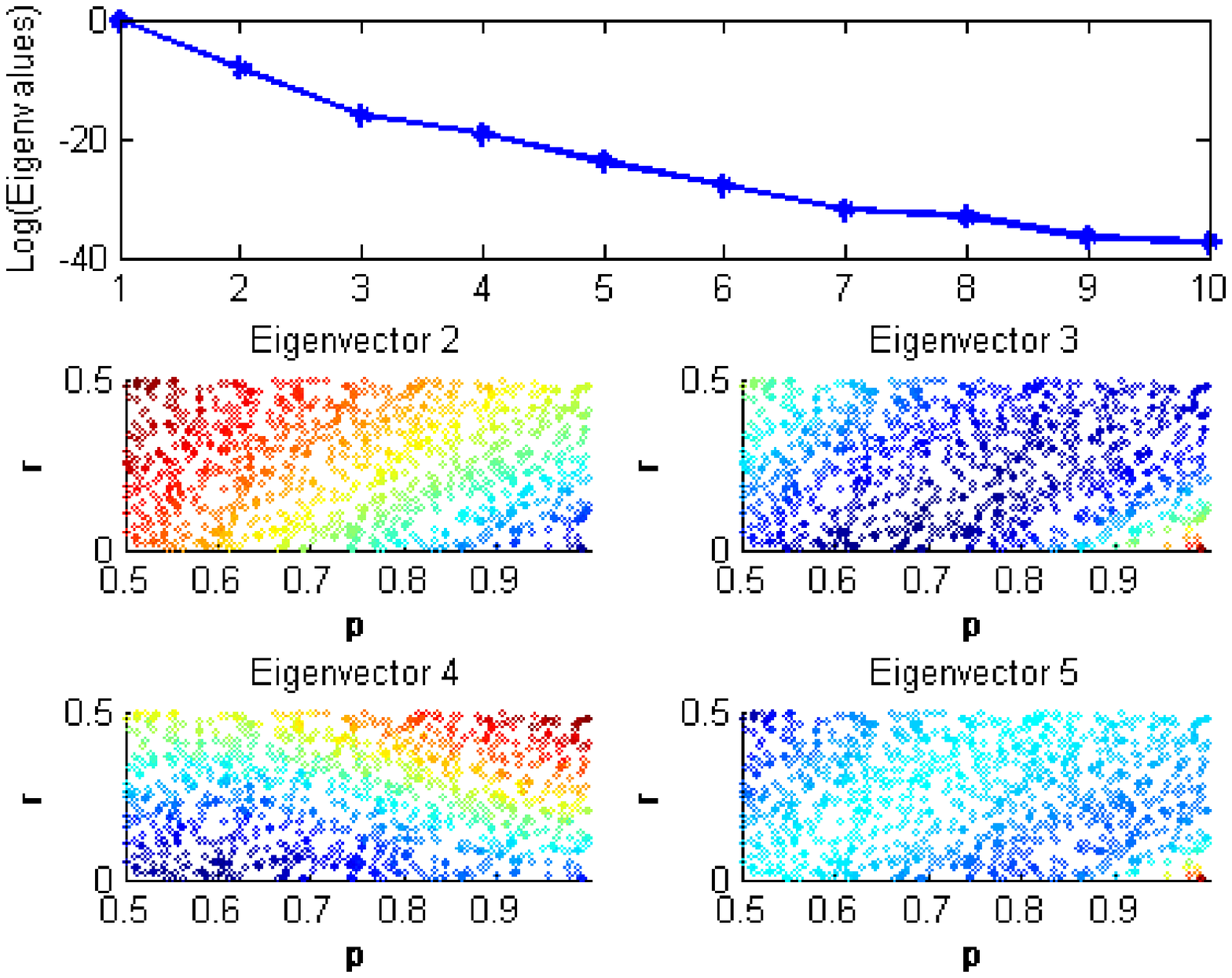}
\caption{\label{fig:CL2} Data mining ensembles of two-parameter Chung-Lu graphs:
The leading eigenvalues of the random walk matrix calculated using our spectral similarity measure are first plotted.
The corresponding first four non-trivial eigenvectors are then illustrated in a way that brings forth their
relation to the construction parameters $p$ and $r$.
In these plots, each graph is denoted as a point.
The $x$ and $y$ coordinates of the point correspond to the parameters $p$ and $r$ used to construct that particular graph.
The graphs are colored based on the magnitude of their components in the eigenvectors of the random walk matrix $A$.
An $\epsilon$ of 10 was used in the diffusion map algorithm.
}
\end{center}
\end{figure*}

\begin{figure*}
\begin{center}
\includegraphics[width=0.81\textwidth]{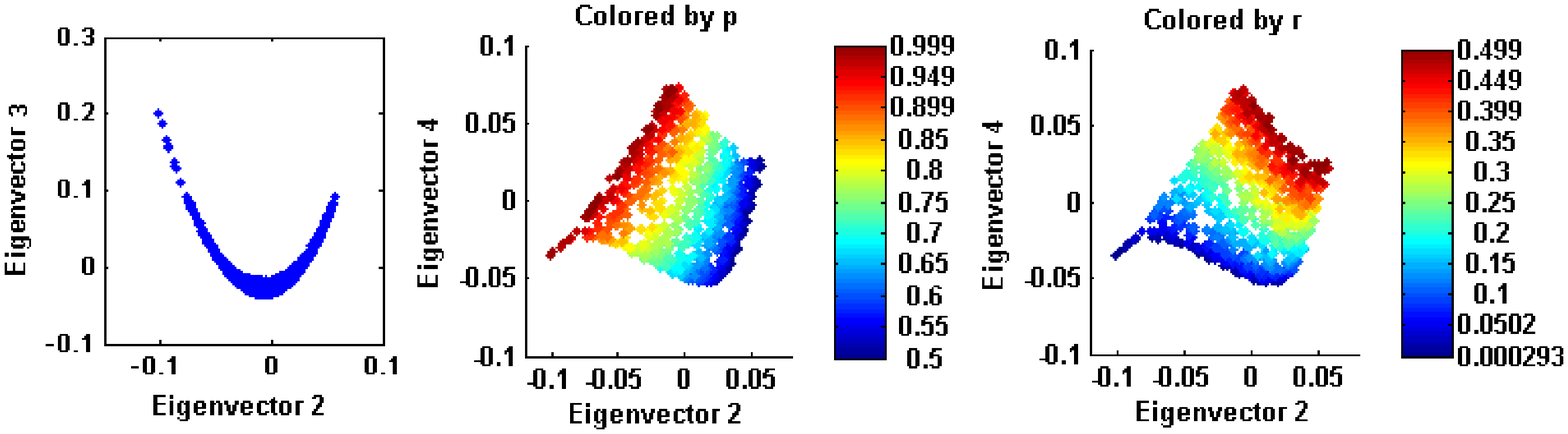}
\caption{\label{fig:CL2ab} Data mining the two-parameter family of Chung-Lu graphs using the spectral similarity metric
leads to an apparent two-dimensional
embedding.
In these plots the $x$ and $y$ coordinates of each point (i.e. of each graph in the dataset)
denote the components of that particular graph in the second and third (resp., fourth) eigenvectors of the random walk matrix
for the left plot (resp., middle and right plots).
Each point is also colored based on the parameter values of
$p$ (middle plot) and $r$ (right plot) used to construct the particular graph.
An $\epsilon$ of 10 was used in the diffusion map algorithm.\footnote{The value of $\epsilon$ is 
representative of the size of the neighborhood around a data point 
over which our metric of choice is considered to be informative. 
As detailed in \cite{Coif08graph}, there is a range of $\epsilon$ values 
which result in meaningful embeddings. Here, for instance, when when $\epsilon$ was reduced from 10 to 0.01 for the analysis
in Fig.~\ref{fig:CL2}, we found that eigenvectors 2 through 14 represented the first principal direction
of the manifold, while eigenvector 15 represented a new direction.
Nevertheless, the low-dimensional manifold recovered using eigenvectors 2 and 15 qualitatively match the
manifolds in Fig.~\ref{fig:CL2ab} plotted using eigenvectors 2 and 4 for the case of $\epsilon = 10$. 
We refer the reader to the work of Clementi et al. \cite{Rohr11determination, Zhen11delineation} 
for additional considerations regarding the choice of $\epsilon$.
}
}
\end{center}
\end{figure*}

We now consider a slightly richer dataset, where
the graphs are constructed using two independent parameters.
The definition of this illustrative family of graphs is based
on the Chung-Lu algorithm \cite{Chun02connected}.
For a graph consisting a $n$ vertices (here $n=100$),
following their original algorithm, we begin by assigning
a weight $w_i$ to  each vertex $i, 1 \leq i \leq n$.
The weights we chose have the two-parameter form  $w_i = np(i/n)^r$.
The probability $P_{ij}$ of existence of the edge between vertices
$i$ and $j$ is given by $P_{ij}=min(B_{ij},1)$, where

\begin{equation}\label{}
B_{ij} = \frac{w_iw_j}{\sum_{k}{w_k}}.
\end{equation}

Once the edge existence probabilities are calculated,
a graph can be constructed by sampling uniform random
numbers between $0$ and $1$ for every pair of vertices $(i,j)$
and placing an edge between them if the random number
is less than $P_{ij}$.
Note that in the original Chung-Lu algorithm $P_{ij}=B_{ij}$.
If the weights are chosen such that $B_{ij} \leq 1, \forall (i,j)$,
then the expected value of the degree of node $i$ will
be equal to the chosen weight values $w_i$.
If any $B_{ij}$ exceeds the value of $1$, this would no longer be the case \cite{Chun02connected}.

The model selected here has $2$ construction parameters: $p$ and $r$.
If $r=0$, the resulting graphs are Erd\"{o}s-R\'{e}nyi
graphs and the parameter $p$ represents the edge density.
When $p=1$ and $r=0$, the resulting graphs are complete.
As $r$ is increased, this procedure creates graphs whose
degree distributions are skewed to the left (long tails towards lower degrees).
Degree distributions resulting from creating graphs with various combinations
of parameters $p$ and $r$ are shown in Fig.~\ref{fig:CL}.

For our illustration, $1000$ graphs were created using
this model with $n=100$ nodes each.
The values of $p$ and $r$ were chosen by uniformly
sampling in the interval $(0.5,1)$ and $(0,0.5)$ respectively.
The diffusion maps algorithm was used on this set of graphs
exactly as described in the first case.
As we will discuss below, the results obtained using the
two similarity measures that we consider in this paper,
while conveying essentially the same qualitative information,
have visible quantitative differences.

The first $10$ eigenvalues of the random walk matrix calculated
using the subgraph approach for evaluating similarities are shown in
the top plot of Fig.~\ref{fig:CL1}.
The first four non-trivial eigenvectors are plotted below.
In these plots, each of the $1000$ graphs is represented as a point
in the $p-r$ two parameter plane.
The colors represent the magnitude of the components of the corresponding
graph data on each of the first $4$ non-trivial eigenvectors.
The gradient of colors in these plots suggest the ``direction"
of each of these eigenvectors in the $p-r$ plane.
However, a more careful inspection of the plots is required to determine
independent subsets of these eigenvectors.
To help explore this, we plot eigenvectors $2$ and $3$ against
each other in Fig.~\ref{fig:CL1a}.
The figure clearly suggests (through its obvious two-dimensionality) that these two eigenvectors are independent
of each other.
Furthermore, when the points in these plots are colored by the two parameters
$p$ and $r$ used to construct the graphs, two independent directions,
a roughly ``left-to-right" for $p$ and a roughly ``top-to-bottom" for $r$,
can be discerned on the $v_2 - v_3$ manifold, Fig.~\ref{fig:CL1a}.
This strongly suggests that the Jacobian of the transformation from $(p,r)$ to $(v_2,v_3)$
is nonsingular on our data.
Thus, these two eigenvectors, obtained solely through our data mining approach,
can equivalently be used to parameterize the set of graphs constructed using the parameters $p$ and $r$.
The components of the fourth eigenvector plotted in
terms of these two leading eigenvectors in Fig.~\ref{fig:CL1b}
are strong evidence that this fourth eigenvector is completely determined by
(is a function of) the second and third ones.
In other words, the fourth eigenvector ``lives in the manifold" created
by the second and third eigenvectors, and hence does not convey
more information about (does not parameterize new directions in) our graph dataset.

We now focus on similar results obtained with the same dataset,
but now using our spectral approach for measuring similarity.
The eigenvalues and eigenvectors of the random walk matrix
obtained by this approach are reported in Fig.~\ref{fig:CL2}.
As before, we plot the leading eigenvectors against each other
in Fig.~\ref{fig:CL2ab}.
The plot of eigenvector $2$ versus eigenvector $3$ appears as a smooth ``almost" curve,
suggesting a strong correlation,
while the plot of eigenvector $3$ versus eigenvector $4$ clearly shows two-dimensionality.
These figures suggest that eigenvectors
$2$ and $3$ parameterize the same direction in
the $p-r$ plane, while eigenvector $4$ parameterizes a second, new
direction in this plane.
Hence, eigenvectors $2$ and $4$ constitute independent directions in the space
of our sample graphs.

Once again, we  have recovered (through data mining) two independent directions in our
sample family of graphs that were originally constructed using two independent parameters.
Although the results obtained using the subgraph and the spectral approaches
in this case are quantitatively different in their details, they are both successful in recovering
two independent coordinates in the space of graphs, apparently isomorphic to those
given as input to the data mining algorithm.
The $2D$ manifold resulting from the subgraph approach appears
better at visually capturing the behavior of the original $p-r$
plane.
The ``quality" of these parameterizations will clearly be affected by the
details used in the data-mining procedure and, in particular, those
affecting the similarity measure evaluation: the number of subgraph
densities kept, the choices for numerical constants such as
$\lambda_i$ in the subgraph approach, $\epsilon$ in diffusion maps, etc.
An obvious criterion in the selection of these method parameters is to make
the Jacobian of the transformation from the ``natural" to the ``data-mining-based"
parameterizations as far from singular as possible.

\subsection{\label{ss:dm} Test case $3$: Graphs from a dynamic graph evolution model}

\begin{figure*}
\begin{center}
\includegraphics[width=0.81\textwidth]{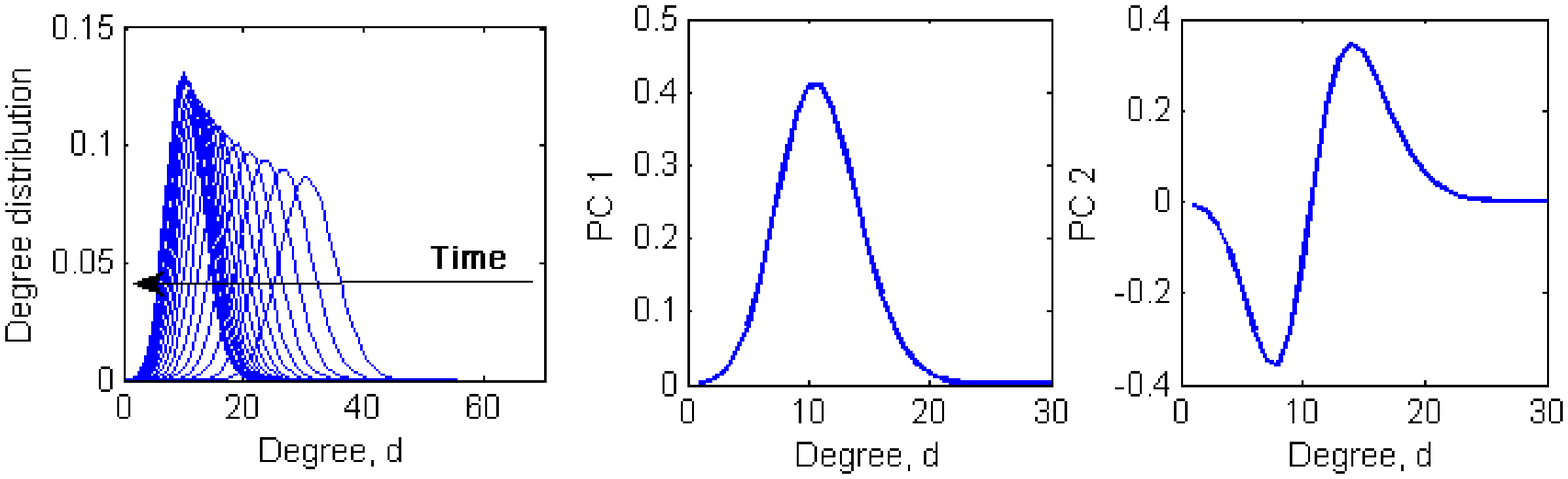}
\caption{\label{fig:K_d3} The evolution of degree distribution over
time from a single initial condition is shown on the left.
The first two principal components (obtained through PCA) of
degree sequences, collected from transients starting from
different initial conditions
are shown on the right.
}
\end{center}
\end{figure*}

\begin{figure*}
\begin{center}
\includegraphics[width=0.81\textwidth]{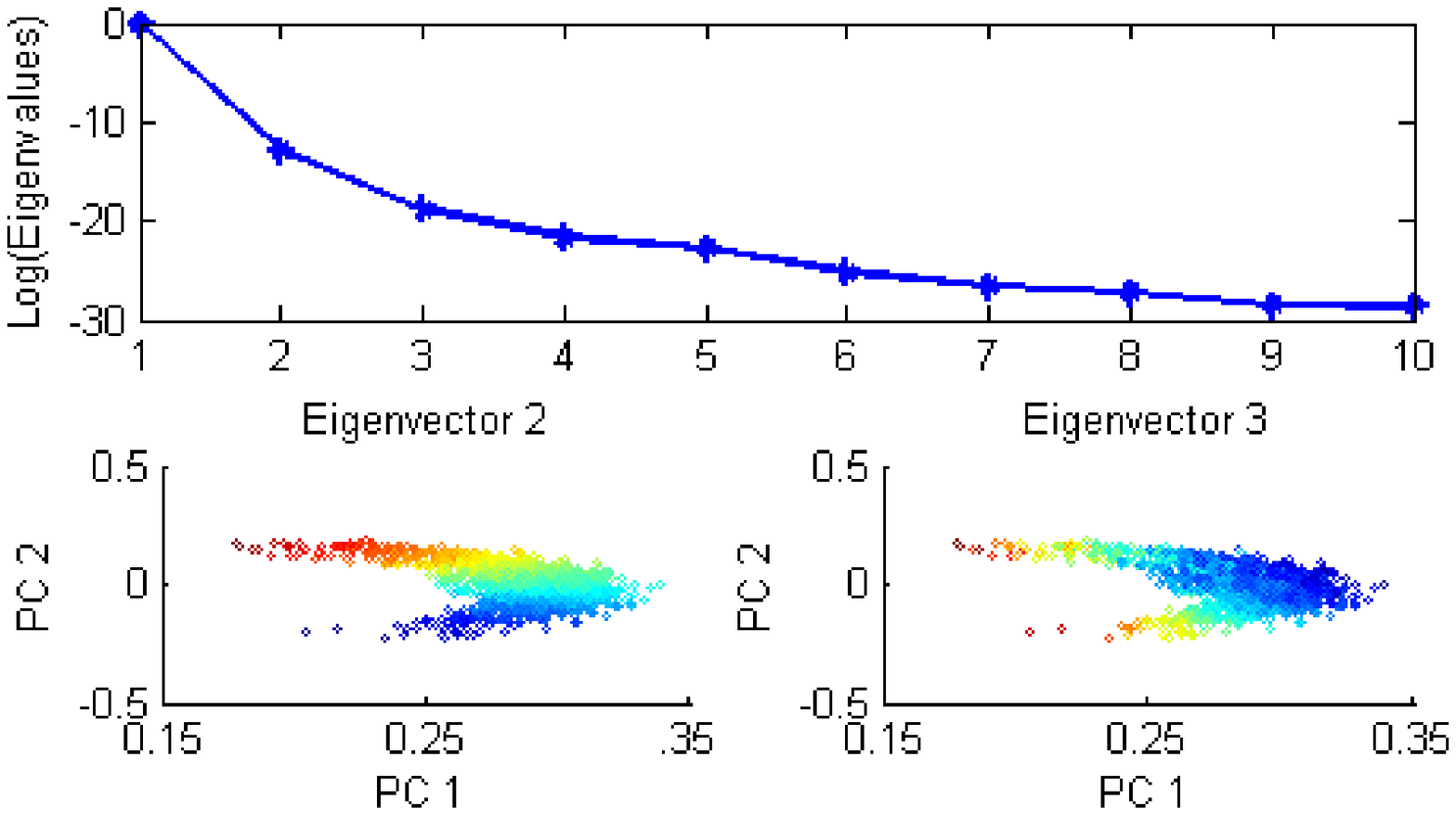}
\caption{\label{fig:K1} Data mining results for graphs collected from the dynamic graph evolution model:
The leading eigenvalues of the random walk matrix calculated using the subgraph similarity measure are plotted.
The corresponding first two non-trivial eigenvectors also are shown. In these plots, each graph datum is denoted as a point.
The $x$ and $y$ coordinates of the point correspond to the parameters PC $1$ and PC $2$ described in the text.
The graphs are colored based on the magnitude of their components in the two leading eigenvectors of the random walk matrix $A$.
An $\epsilon$ of 10 was used in the diffusion map algorithm.
}
\end{center}
\end{figure*}

\begin{figure*}
\begin{center}
\includegraphics[width=0.81\textwidth]{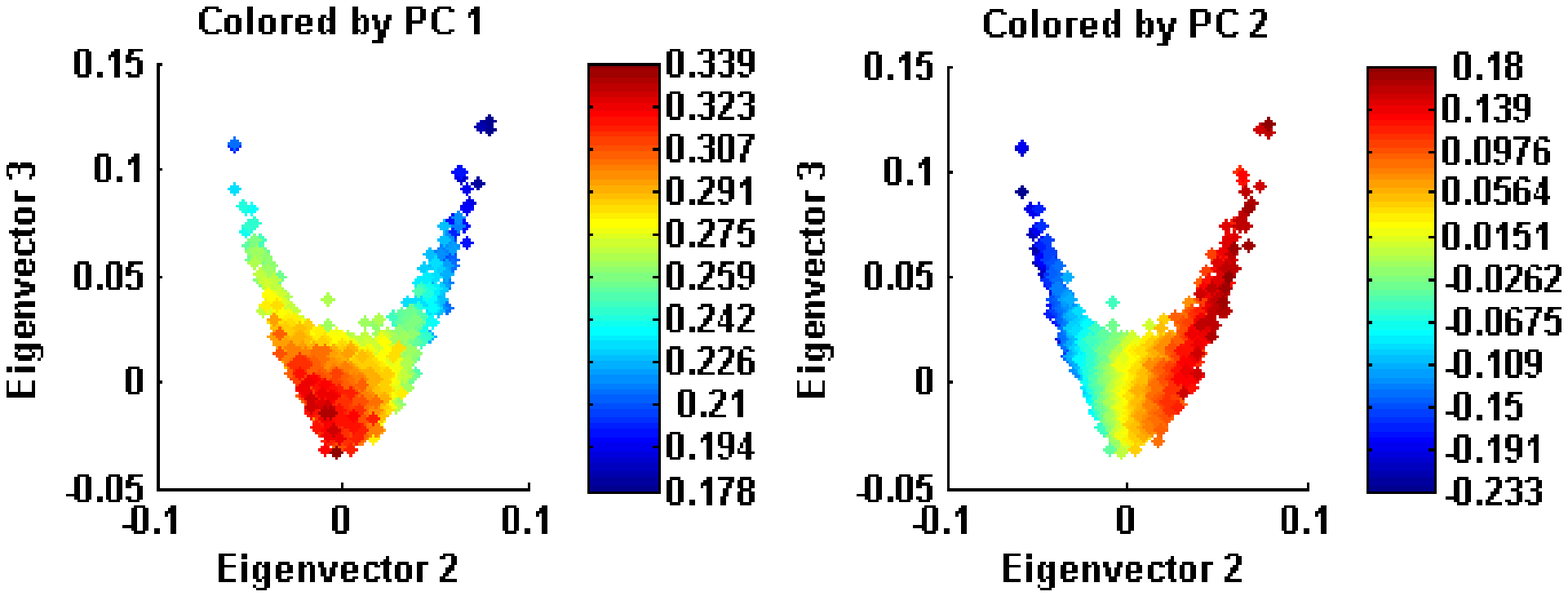}
\caption{\label{fig:K1a} The graph data collected from the dynamic model transients are
alternatively plotted in the embedding plane of the two leading diffusion map eigenvectors shown in Fig.~\ref{fig:K1}.
The points corresponding to the different graphs are colored based on the component of their degree
distributions on the first two degree-distribution principal components.
}
\end{center}
\end{figure*}

\begin{figure*}
\begin{center}
\includegraphics[width=0.81\textwidth]{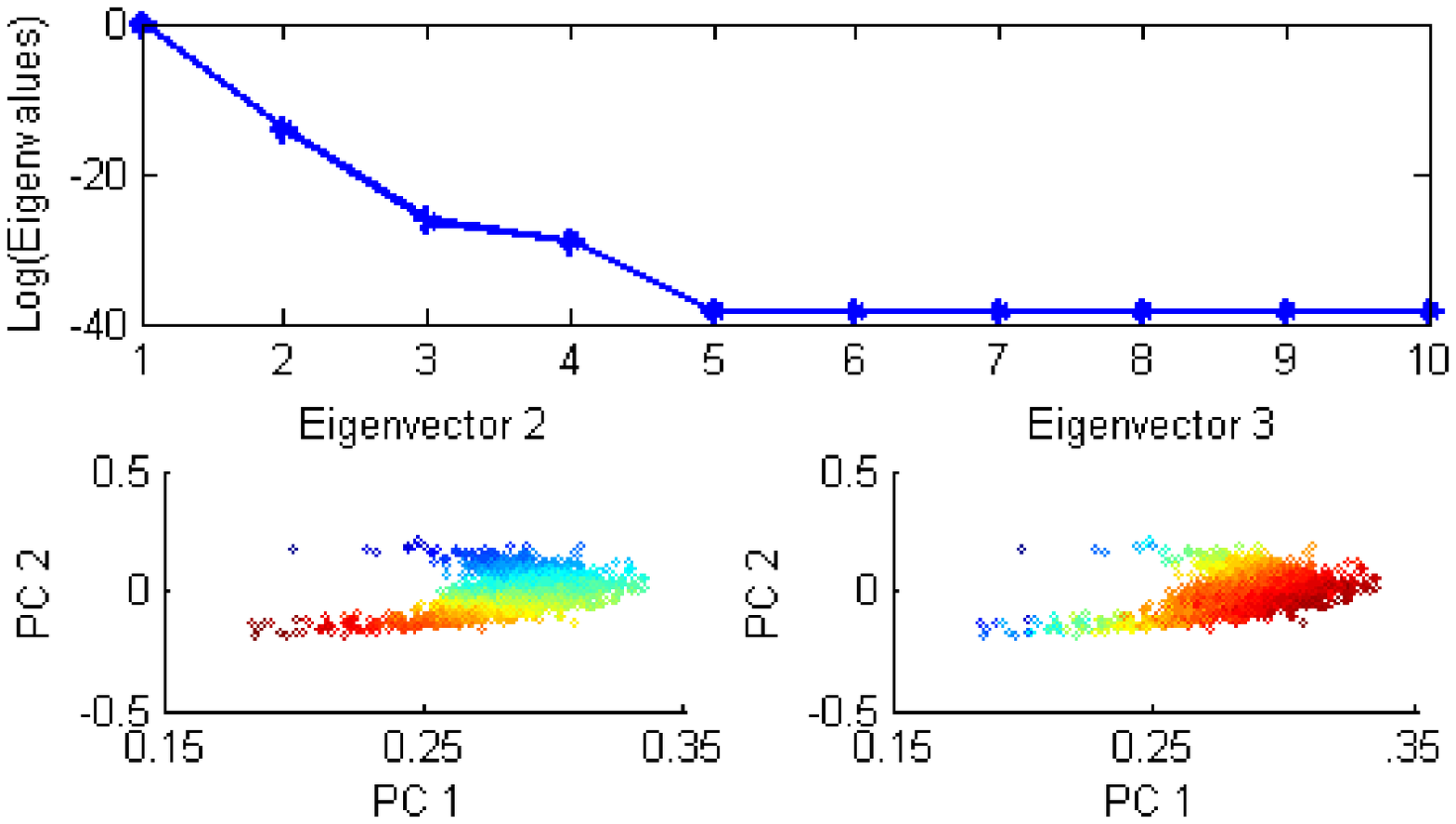}
\caption{\label{fig:K2} Data mining results for graphs collected from the dynamic graph evolution model:
The leading eigenvalues of the random walk matrix calculated using our spectral similarity measure are plotted.
The corresponding first two non-trivial eigenvectors also are shown. In these plots, each graph datum is denoted as a point.
The $x$ and $y$ coordinates of the point correspond to the parameters PC $1$ and PC $2$ described in the text.
The graphs are colored based on the magnitude of their components in the two leading eigenvectors of the random walk matrix $A$.
An $\epsilon$ of 50 was used in the diffusion map algorithm.
}
\end{center}
\end{figure*}

\begin{figure*}
\begin{center}
\includegraphics[width=0.81\textwidth]{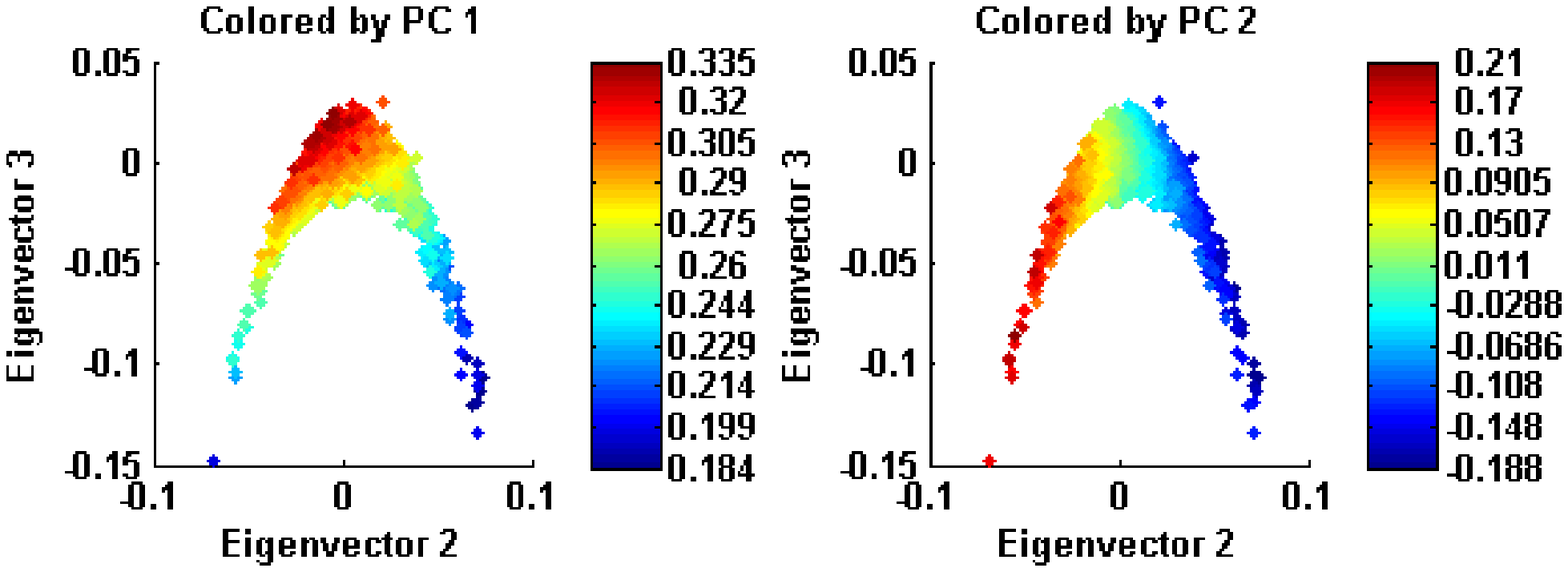}
\caption{\label{fig:K2a} The graph data collected from the dynamic model transients are
alternatively plotted in the embedding plane of the two leading diffusion map eigenvectors shown in Fig.~\ref{fig:K2}.
The points corresponding to the different graphs are colored based on the component of their degree
distributions on the first two degree-distribution principal components.
}
\end{center}
\end{figure*}

In the two examples given above, the dataset graphs were created
using prescribed rules of a graph-generation model characterized  by one or more model parameters.
We now consider a case where the graph dataset comes from sampling
snaphots of a dynamic process that involves the evolution of graphs.
The process sampled could be one occurring in nature/society (for instance, evolution
of a social network) or one arising from a dynamic model
with prescribed (deterministic or even stochastic) evolution rules.
We consider the latter case for illustration, exploiting a simple
model of a random evolution of networks \cite{Bold12Equation-Free};
our process is sampled at regular time intervals.
A brief description of the model is as follows:
Starting from an initial graph, the model rules update the
graph structure at every time step by repeatedly sequentially applying the
following two operations:

\begin{enumerate}
\item A pair of nodes selected at random are connected by an edge if they
are not already connected to each other.
\item An edge chosen uniformly at random is removed with probability $r$ (here, $r=0.1$).
\end{enumerate}

The details of the model behavior are discussed in \cite{Bold12Equation-Free}.
Here, we will focus only on those characteristics of the model evolution
necessary to rationalize the results of data mining.
For this particular graph model, it is known that the (expected) degree distribution
evolves smoothly in time as shown in the left plot of Fig.~\ref{fig:K_d3}.
Furthermore, it is also known that the evolution of degrees is decoupled from
(and occurs at slower time scales than) the evolution of all higher order properties of the graph
such as triangles, degree-degree correlations etc.
Thus, as argued in \cite{Bold12Equation-Free}, the dynamic evolution of graphs according to this
model can be usefully described in terms of degrees or degree distributions only.
The temporal evolution starting from several different initial graphs was recorded,
and a principal component analysis of a collection of snapshots of degree sequences that
arose during these simulations was performed.
The two leading principal components
are plotted in Fig.~\ref{fig:K_d3}.
The first principal component (PCA), labeled PC $1$, is, in effect, the steady state degree distribution
while the second principal component PC $2$ is the direction along which
the degree distribution decays the slowest towards stationarity.
Since it has been claimed that the degree distribution is the most significant observable in
this model, PC $1$ and PC $2$ constitute good variables through which one can track the evolution of
the graphs over time.
In fact, as shown in  \cite{Bold12Equation-Free}, one can write explicit
Fokker-Planck equations for the evolution of the distribution of
(appropriately shifted and scaled) degrees.
The eigenfunctions of the corresponding eigenvalue problem are Hermite polynomials,
the first two of which have the qualitative forms, $exp(-x^2)$ and $x.exp(-x^2)$;
this functional dependence is clearly noticeable in $PC1$ and $PC2$ respectively.

We now ignore our knowledge of explicit models of the dynamical process,
and focus on the sampled graph sequences created by the process.
Our goal  is to use diffusion maps to discover -using only the graph snapshot data-
good variables through which to characterize the data.
These  variables should then also be useful in describing the dynamics of the
evolution process -- we anticipate that the data-mining variables and the variables
PC $1$ and PC $2$ would embody similar information about the process.
As before, we use both the subgraph and our spectral similarity measures.
The eigenvalues and the first two non-trivial eigenvectors of the random walk
matrix in diffusion maps are shown for each similarity measure
in Figs.~\ref{fig:K1} and \ref{fig:K2} respectively.
The plots are constructed so that the plotted points (each corresponding to a
sampled graph) lie in the plane of the principal components PC $1$
and PC $2$; the points are colored by the corresponding eigenvector components, for comparison.
In both cases, the results show that the second and third eigenvectors have the most variation
(as indicated by the gradient of colors) in the directions of PC $2$ and PC $1$ resp.
This suggests that the first two non-trivial eigenvectors are roughly one-to-one with
PC $2$ and PC $1$ respectively.
Conversely, we can plot the graphs in the space of the diffusion map eigenvectors (the new embedding)
and color them based on PC $1$ and PC $2$ as shown in Figs.~\ref{fig:K1a} and \ref{fig:K2a}
(one for each of our similarity measures).
These two figures convey again the one-to-one nature of the transformation between
Principal Component based and diffusion map based parameterization of the dynamic
graph evolution data - yet the diffusion map based parameterization {\bf did not} use prior knowledge about
significant observables of the process.
In the example considered here, the fact that theoretical results were available
was used to validate our data mining, providing a reference with which to compare.
The implication is that, even in problems where such theoretical results are not available,
data mining can be used to gain an understanding about the primary driving factors
in the dynamics of the system - provided that a relatively small number of
such factors can be used to effectively model the process.

\section{\label{sec:conc} Conclusions}

We studied the problem of data mining in cases where
the data points occur in the form of graphs.
The main obstacle in applying established data mining algorithms to
such cases is the definition of good measures quantifying the similarity
between pairs of graphs.
We discussed two common sense approaches to tackle this problem: {\em the subgraph
method}, which compares the local structures in the graphs, and {\em a spectral
method}, which is based on defining diffusion processes on the graphs.
While alternate definitions of similarity metrics than the ones discussed in
this paper are eminently possible, and probably very useful,
the purpose of this paper is to demonstrate the
usefulness of data mining in the context of graph data using a few illustrative
examples for which the parameterizations obtained through our approach
could be compared with known results.
Certain remarks regarding the similarity measures
used in this paper should however be made.
The subgraph approach to evaluate similarity is much more computationally expensive, compared to our
spectral approach, especially when larger sized subgraphs must be included to get
useful results.
(For example, there are $6$ connected subgraphs of size $4$,
while there are $21$ subgraphs of size $5$. Searching for larger subgraphs becomes computationally
increasingly more
expensive).
Both approaches require us to tune certain method parameters associated with the
definitions of the similarity metric and the data mining approach.
For the diffusion map algorithm, one has to choose a suitable size of
neighborhood ($\epsilon$).
In addition, the spectral approach requires one to define the weighting function, $\mu(k)$
(and also make assumptions about the vectors $p$ and $q$).
This can be thought as analogous to selecting suitable normalizations
in defining the subgraph densities in the subgraph approach.
Such tuning considerations become especially crucial when one is confronted
with data from a completely new problem, where intuition cannot be used to guide
the selection of model parameters.
Considering the trade-offs mentioned above, it might be reasonable to use
the subgraph density approach to find similarities between graphs initially for new problems,
and to subsequently ``tune" spectral decomposition algorithm, which can then be used for faster computations.
%

%
We used three sample sets of graph data in this work.
The first example was a collection of Erd\"{o}s-R\'{e}nyi random graphs
with varying parameters.
We also considered the case of graphs obtained from a simple two-parameter
family of graphs motivated by the Chung-Lu algorithm.
Both examples considered graphs created from a fixed model.
As a third example, we used a collection of graphs from a dynamic model.
In all these examples, we used the data mining approach with two different
approaches for measuring similarities to extract good characterizations of the
graph datasets with identical size graphs, and compared them to known parameterizations.
A weighted version of the graph size can be used as part
of the selected similarity measure; so that the measures we used
can be extended to datasets with varying graph sizes.

An important motivation for our work is the desire to use the data mining-induced observables
as variables in the equation-free modeling of graph evolution (\cite{Kevr03Equation-free,Kevr04Equation-free:},
see \cite{Bold12Equation-Free,Tsou12coarse-graining} for graph-related applications).
Even if explicit models for the coarse-grained evolution of graphs may not be available/easy to derive,
the idea is that one can still use well-designed brief bursts of detailed network evolution simulations
to effectively perform coarse-scale computations.
In this process -and the associated algorithms, like coarse projective integration- the ability to fluently
translate from fine (detailed network) to coarse-grained (reduced, only important statistics) descriptions
lies at the heart of the approach.
Several good attempts for implementing the reverse step of this transformation
({\em lifting} from coarse statistics to detailed networks consistent with them)
can be found in the literature \cite{Have55remark,Doro02how,Holm02growing,Serr05tuning,Goun11generation};
using data-mining inspired observables in such equation-free coarse-graining is a significant challenge,
one that we are currently exploring.

\begin{acknowledgments}
This work was partially supported by the US Department of Energy (DE-FG02-10ER26024 and DE-FG02-09ER25877).
The authors are grateful for extended discussions with Prof. Amit Singer of the
Mathematics Department/PACM in Princeton, in whose data-mining class this project
originated.
\end{acknowledgments}

\bibliographystyle{apsrev4-1}
\bibliography{All}

\end{document}